\documentclass[preprint]{aastex}
\begin{document}

\title{Suites of dwarfs around nearby giant galaxies}
\author{Igor D. Karachentsev \altaffilmark{1,2}, Elena I. Kaisina \altaffilmark{1} and Dmitry I. Makarov \altaffilmark{1,2}}
\affil{$^1$ Special Astrophysical Observatory RAS,  Nizhnij Arkhyz,    Karachai-Cherkessian Republic,
    Russia 369167}
\affil{$^2$ Leibniz-Institut f\"{u}r Astrophysik (AIP), An der Sternwarte 16, D
-14482, Potsdam, Germany}
\email{ikar@sao.ru,kei@sao.ru,dim@sao.ru}

\begin{abstract}

Updated Nearby Galaxy Catalog (=UNGC) contains the most
comprehensive summary of distances, radial velocities and 
luminosities for eight hundred galaxies located within 11 Mpc 
from us. The highest density of observables in UNGC makes this
sample indispensable for checking results of N-body
simulations of cosmic structures on a $\sim1$ Mpc scale.

Environment of each galaxy in UNGC was characterized by a 
tidal index $\Theta_1$ depending on the separation and mass 
of the galaxy's Main Disturber (=MD). We ascribed the UNGC 
galaxies with a common MD to its suite, and ranked suite 
members according to their $\Theta_1$. All suite members with 
positive $\Theta_1$ are assumed to be physical companions of the MD.
About 58\% of the sample are members of physical groups. 

The distribution of suites by the number of members, n, follows
to a relation $N(n) \sim n^{-2}$. The twenty most populated suites
contain 468 galaxies, i.e. 59\% of the UNGC sample. The fraction of MDs
among the brightest galaxies is almost 100\% and drops to 50\% at
$M_B = -18^m$.

We discuss various properties of MDs, as well as galaxies
belonging to their suites. The suite abundance practically does
not depend on morphological type, linear diameter or hydrogen
mass of MD, revealing the tightest correlation with the MD
dynamical mass. Dwarf galaxies around MDs exhibit well-known
segregation effects: the population of outskirts has later
morphological types, richer HI-contents and higher rates of star
formation activity. Nevertheless, there are some intriguing cases
when dwarf spheroidal galaxies occur at the far periphery of the
suites, as well as some late-type dwarfs residing close to MDs.

Comparing simulation results with galaxy groups, most studies
assume the Local Group is fairly typical. However, we recognize
that the nearby groups significantly differ from each other and
there is a considerable variation in their properties. The suites
of companions around the Milky Way and M 31, consisting the Local
Group, do not look as a quite typical nearby group.

The multiplicity of nearby groups on number of their physical 
members can be described by the Hirsh-like index $h_g = 9$, 
indicating that the Local Volume contains 9 groups with 
populations exceeding 9 companions to their MDs.

 \end{abstract}

\section{Introduction}

The standard LCDM cosmological model  with cold dark matter and
dark energy efficiently explains the observed properties of the
universe on large scales (Klypin et al. 2003). 
The modern cosmological N-body simulations have resolutions good enough 
to investigate structures with size about or better than 1 Mpc and 
with individual halos about $10^7$ solar masses (Klypin et al. 2011, Kitaura et al. 2012).
However, our advances in the
matching the simulation results with the observational data on
such small scales still look very modest. One reason for this is a
limited database on the distances even to the nearest galaxies.

Over the last 10--15 years, mass measurements of distances to the
nearby galaxies have been undertaken by several observational
teams, relying on the unique resolution of the Hubble Space
Telescope (HST). Use of a Tip of the Red Giant Branch (TRGB) stars 
as a ``standard candle'' (Lee et al. 1993) allows to determine 
the distances for more than 300 of the most nearby galaxies 
with an error of $\sim10$ \%.
The
first summary of the new and old distance estimates  was
presented in the catalog of galaxies of the Local Volume
(Karachentsev et al. 2004), which contains data on 450 galaxies
in a sphere of 10 Mpc radius around the Milky Way. Later on, the
distance estimates and other integral parameters of nearby
galaxies have been accumulated in the Extragalactic Distance
Database (http://edd.ifa.hawaii.edu) by Tully et al. (2008) and
Database on the Local Volume Galaxies (http://www.sao.ru/lv/lvgdb) by
Kaisina et al. (2012). The Updated Nearby Galaxy Catalog 
(Karachentsev et al. 2013 = UNGC) contains the most complete
summary of various observable characteristics for $\sim800$
galaxies located within 11 Mpc. The UNGC catalog is currently the
most representative and homogeneous sample of neighboring galaxies, most of
which have known linear separations, luminosities and line-of-sight
velocities. 
Unlike most catalogs which are limited by flux, this sample is restricted by distance. 
It makes UNGC the most suitable for comparison with N-body simulations on the small scales $\sim$(0.1--10) Mpc.

\section{Environment of nearby giant galaxies}

For each of the 869 galaxies in the UNGC catalog (Karachentsev et al. 2013) 
we determined its ``tidal index'' (Karachentsev and Makarov, 1999)
$$\Theta_1=\max[\log(L_n/D^3_n)]+C, \,\,\,\,\, n=1,2,\ldots N \eqno(1) $$
where $L_n$ is the $K$-band luminosity of the neighboring galaxy, and $D_n$ 
is its spatial separation from the considered galaxy. Ranking the
surrounding galaxies by the value of their tidal force
$F_n\sim L_n/D^3_n$, we are looking for the most
significant, influential neighbor, which is designated as the
Main Disturber (MD).
We assume that the total mass of the galaxy is proportional to its 
luminosity in $K$-band, and that the mass-to-light ratio does not 
depend on the luminosity and morphology. The constant
$C=-10.96$ in (1) has been chosen so that the galaxy with $\Theta_1=0$
locates on the ``zero velocity sphere'' relative to its MD. In other
words, the galaxy with  $\Theta_1>0$ is considered to be
causally connected with its MD, since the crossing time for this
pair is shorter than the age of the universe  $H_0^{-1}$, where
$H_0=72$ km s$^{-1}$ Mpc$^{-1}$ is the Hubble parameter.
Accordingly, a galaxy with a negative $\Theta_1$ should be
considered as  physically not bound with its neighbors. Such
objects are usually referred to as the ``field'' galaxies.
Evidently, this approach is only justified for the close volume
where all the fairly massive galaxies are already discovered and
their distances have been measured.

In the Tables 1 and 2 of the UNGC catalog (Karachentsev et al. 2013)
we have presented the observing and physical characteristics 
of 869 of the Local Volume galaxies, 
taking into account corrections for external and internal extinction. 
We have
excluded from this sample 75 galaxies with distance estimates of
$D>11.0$~Mpc and united all the remaining objects in associations
with their common Main Disturber. We call the set
of galaxies with one common Main Disturber as the MD
``suite''. Within each suite, its members were ranked by highest
tidal index $\Theta_1$. A sub-sample of members of the suite with
$\Theta_1\geq 0$ we determine as a physical group, where
the MD is the dominant galaxy by mass. In almost all the cases,
the groups of galaxies formed this way matched with the list of
nearby groups by Karachentsev (2005).

The suites around the MDs themselves were ranked according to the
number of suite members $n_s$ from the maximum of   $n_s=53$ for
the suite around M81 to  $n_s=1$. The sample of galaxies of
the Local Volume  reorganized this way is presented in Table~1,
the full version of which is available at the LVG page on
the website of
the Special Astrophysical Observatory of the Russian Academy of the Sciences
(http://www.sao.ru/lv/lvgdb).

The table columns contain the following data:

(1)  the name of the galaxy;

(2)  linear diameter of the galaxy in kpc, determined at
Holmberg's isophote (26.5 mag/square arcsec);

(3)  absolute magnitude of the galaxy in the B-band corrected for
extinction;

(4)  logarithm of the stellar mass in solar units;

(5)  logarithm of the indicative (dynamic) mass within the
Holmberg diameter, $\log(M_{26}/M_{\odot})=2\log V_m +\log a_{26}
+\log D +4.52,$ where the rotation velocity $V_m$ is expressed in
km s$^{-1}$, the Holmberg diameter $a_{26}$ --- in angular
minutes, and the distance $D$ --- in Mpc;

(6) logarithm of the hydrogen mass in solar units;

(7) tidal index $\Theta_1$;

(8) Main Disturber's name;

(9) distance to the galaxy in Mpc;

(10) line-of-sight velocity of the galaxy (in km s$^{-1}$)
relative to the velocity of MD;

(11) number of members in the suite of MD, to which the
galaxy belongs.

The distribution of the number of suites around the MDs by the
number of their members is demonstrated in Fig.~1 in the 
logarithmic scale. Open circles in the figure correspond to the total
number of galaxies in the suite with any tidal indices. The filled
circles show the number of bound companions, satisfying
$\Theta_1\geq 0$. Standard errors   $\sqrt{N}$ are depicted by
vertical bars. In general, the distribution of suites by the
number of galaxies in them is  represented quite well by the
power law  $N(n)\propto n^{-2}$, which is described in the figure
by the straight line.

Among the 794 galaxies of the Local Volume 457 galaxies or 58\%
have $\Theta_1 \geq 0$ values. In other words, they are the members of
physical groups of different multiplicity. It should be noted
that according to Makarov \& Karachentsev (2011), for $\sim11000$
galaxies of the Local Universe located within the sphere of
$D\simeq 50$ Mpc radius, the relative number of galaxies in
groups is 52\%. Thus the abundance of galaxy group members in small
and large volumes is almost the same.  The agreement of
these quantities can be considered as some evidence of
representativeness of the Local Volume in terms of structure and
dynamics of galaxy systems.

The data in Table~1 show that most of the galaxies in the Local
Volume are concentrated in  suites around a small number of the
most massive galaxies. Thus, only 20 most populated suites
contain 468 galaxies, i.e. 59\% of the total population of the
Local Volume.\footnote{We have not included in this list a suite
of 12 galaxies around NGC~4414, which lies outside the LV at a
distance of 18 Mpc, neither the suite of 10 galaxies around
NGC~1291, the distance to which  is very uncertain.} Some
parameters of these 20 structures and properties of their main
galaxies are presented in Table~2 with columns containing: (1)
the abbreviated name of the main galaxy (MD); (2)  the distance
to the MD in Mpc, by which the list of suites is ordered; (3) the
total number of galaxies in the MD suite, including the field
objects; (4) the number of physical group members with
$\Theta_1\geq0$; (5) the number of ``bright'' bound companions
of the main galaxy with absolute magnitudes
 $M_B$ brighter than $-11.0^m$; (6)  the absolute
magnitude of the main galaxy; (7, 8)
  its stellar as well as dynamical masses within the Holmberg
diameter in solar masses; (9) linear Holmberg diameter of MD in kpc;
(10,11) hydrogen mass and morphological type of the main galaxy by
 de Vaucouleurs classification; (12 - 14) tidal indices,
characterizing the MD environment: $\Theta_1$ ---
tidal index determined by the most significant neighbor; $\Theta_5$
--- tidal index determined by the total contribution of the five most
significant neighbors,  $\Theta_ 5=\log(\sum^{5}_{n=1}
M_n/D^3_n)+C$; and $\Theta_j=\log(j_*[1 \rm{Mpc}]/j_{*,global})$
--- logarithm of the mean density of stellar mass around
the galaxy (excluding the galaxy itself) within a 1 Mpc radius,
expressed in units of the global mean density
$4.28\times 10^8 M_{\odot}\cdot$Mpc$^{-3}$ (Jones et al. 2006).

The distribution of members of the 20 most populated suites by the
tidal index $\Theta_1$ is shown in Fig.~2. As follows from it,
about 60\% of members of these suites have $\Theta_1\geq 0$,
i.e., are physically bound with the main galaxy. It should be
noted, however, that not all the galaxies of the Local Volume
have their distances measured with high accuracy. Therefore, the
 $\Theta_1=0\pm0.5$ boundary strip may contain galaxies of different
status: both the group members and field galaxies.

We noted above that more than a half of the total population of
the Local Volume is located in the field of   gravitational
influence of only 20 giant galaxies. Figure~3 represents the
distribution of galaxies of the Local Volume by the absolute
$B$-band magnitude. The inset picture shows what fraction of 
the Main Disturbers as function of the absolute magnitude.

The relative number of MDs among the brightest galaxies is close
to 100\%. As might be expected, the fraction of MDs decreases
towards the low-luminosity galaxies, dropping below 50\% at
$M_B\simeq-18.0^m$. A similar pattern was noted by Wang \& White
(2012) according to the data on SDSS survey. MDs are also presented on the
faint end of the luminosity function. They can be conditionally
divided into two categories: a) dwarf companions located close to
a giant galaxy (an example is a dwarf spheroidal system SagdSph,
tumbledown by the tidal influence of the Milky Way), b) tight
pairs of dwarf galaxies, for example, UGCA~319+DDO~161,
KK~78+DDO~64, KK~65+DDO~47, where each component of the pair is a
MD for the second component. The list of similar isolated
multiple dwarf galaxies in the volume of
 $\sim50$ Mpc radius was compiled by Makarov \& Uklein (2012).

It should be stressed that the considered sample of nearby galaxies
suffers with different selection effects. Clearly, these are very
complex and variable due to the heterogeneous nature of many of
the surveys that contribute to the UNGC catalog. For instance, 
there are a luminosity bias with distance, a HI-bias over the
sky because of the declination horizon and limited angular resolution
for radio telescopes, etc. In particular, blind HI-surveys, like HIPASS
and ALFALFA, are efficient to reveal gas-rich irregular dwarfs in
the Zone of Avoidance, which are practically invisible in optical
surveys, but the radio surveys are nearly insensitive to detect
dwarf spheroidal objects.

\section{Some properties of the Main Disturbers}

Returning to the Table~2 data, let us note some features of the
main galaxies in the suites, which foster the presence  of a
large number of companions around  MDs. Four panels of Fig.~4
show the dependence of the number of physical members of the suite
(i.e. members of group), $n_g$, on the stellar and dynamical mass 
of the MD, as well as its linear diameter and hydrogen mass. As one 
can see, the most obvious relationship occurs for the dynamic mass 
of the main galaxy $M_{26}$, which was earlier noticed by 
Karachentsev \& Kasparova (2005). It should be noted, however, 
that due to the selectivity by  luminosity, the 
suites of nearby MDs look more populated than the suites of their 
distant counterparts. To reduce the selectivity  effect with
distance, we have excluded from our analysis the dwarf galaxies
with absolute magnitudes   $M_B>-11.0^m$. The reduced number of
bright physical companions is indicated in Table~2 as $n_b$.

Considering each parameter in Table~2 as a feature that may
affect the number of members of the suite, we calculated the
correlation coefficients of these parameters with the total
number of galaxies in the suite $n_s$, the number of physical
members $n_g$ and the number of bright physical companions
$n_b$. The results are shown in Table~3.

If we assume the correlation coefficients larger than
0.25 by modulus to be significant, then the data in Table~3 leads
to the following conclusions. a) Linear diameter of the main
galaxy, its hydrogen mass and morphological type
have practically no effect on the population of a suite. 
b) The total number of members of the suite $n_s$ and the
number of bound companions  $n_g$ show a positive correlation
with the luminosity of the main galaxy, its dynamical mass
$M_{26}$, and all three tidal indices $\Theta_1, \Theta_5,
\Theta_j$;  however, the presence of a significant correlation
between $n_s$ and $n_g$ with distance  indicates the effect of
observational selection as the cause of  listed correlations. c)
For  bright physical group members, $n_b$, the correlation with
distance $D$ virtually disappears. The number of $n_b$ is
significantly influenced by the value of stellar dynamic mass of
the MD, and by the contrast of stellar density of the environment, 
$\Theta_j$. But the last circumstance is almost trivial,
since it is the abundance of MD's companions that determines the
density contrast $\Theta_j$. The above trends may shed some light
on the conditions of formation and evolution of massive galactic
halos surrounded by small sub-haloes.

\section{Properties of galaxies in  MD suites}

We know that the groups and clusters of galaxies reveal
segregation effects along the radius by the luminosity,
morphological type and other characteristics. The tidal index
$\Theta_1$ is an indicator of distance of the suite member from
its main galaxy, normalized to the MD mass. This allows to rank
the members of different suites by $\Theta_1$, 
to form a synthetic unified suite.

Figure~5 presents the distribution on several parameters of
galaxies in the 20 most populated suites (Table 2) along the
$\Theta_1$ scale. On its left top panel,  absolute magnitude of
galaxies of the synthetic  suite is clearly correlated with
$\Theta_1$. However, exclusion of galaxies fainter than
  $-11.0^m$ (above the dashed horizontal line), mainly found 
in the vicinity of the Milky Way, Andromeda and M81, makes this 
correlation insignificant. The top right panel of the 
figure shows the hydrogen-to-stellar mass ratio as a function of 
$\Theta_1$. The open circles mark the objects where only the
upper limit of the HI flux is estimated. Despite a large
dispersion of $M_{HI}/M_*$ ratios, its mean value systematically
decreases from the field galaxies towards members of groups.
This known effect is usually explained by sweeping-out
the gas of dwarf galaxies in groups as they pass through the dense
halo regions of a massive galaxy (Slater \& Bell 2013). Note,
however, that among the field galaxies with
  $\Theta_1<0$ there are
objects with low hydrogen abundances per stellar mass unit. To
explain these cases, we need to employ some other mechanisms of
gas loss by dwarf galaxies, for example, the ``cosmic web
stripping'' (Benotez-Llambay et al. 2013).

The lower left panel reproduces the specific star formation rate
(SFR) in the galaxies of synthetic suite as a function of
$\Theta_1$. The $SFR$ was estimated by the $H\alpha$ flux and
ultraviolet $FUV$ flux measured with GALEX. Empty symbols 
correspond to the upper limit of the
$H\alpha$ and $FUV$ fluxes. The smallest scatter in $SFR/M_*$
occurs in the galaxies belonging to outskirts of
the suites. With the growth of $\Theta_1$ there are many
cases of depressed star formation. Just like for the hydrogen-to-stellar mass ratio,
$M_{HI}/M_*$, the decrease of a specific star formation rate, $SFR/M_*$, 
in densest regions is apparently caused by the effect of
sweeping gas out from the shallow potential well of the dwarf
galaxies.

The bottom right panel of Fig.~5 shows the distribution of
galaxies of the synthetic suite by morphological types in de
Vaucouleurs classification at different $\Theta_1$. 
Again, the gas-rich late-type dwarf
galaxies, T = 10, 9 (=Ir, Im, BCD), prevail in the low-density
regions with  $\Theta_1<0$, while the early-type objects, T$<0$
(=E, S0, dSph), are found mainly in the dense central parts of the
suites.
Note that in this panel there are three objects marked with the 
type T = 11. We have classified in this category the intergalactic
HI-clouds without any signs of stellar population. The fact that
two of them have the $\Theta_1>0$ values is likely determined by
the selectivity effect: in the regions of nearby groups, the
HI-surveys are as a rule performed to a deeper extent than in the
vast areas between the groups.

Despite the presence of a quite evident morphological segregation
of galaxies along the radius of the groups, the lower left corner
of the \{$T\propto\Theta_1$\} diagram hosts a number of galaxies
with the characteristics:  T$<0, \Theta_1\leq 0$. These galaxies
can be critical when testing
different scenarios of  formation of early-type galaxies. Twelve
of them are shown in Table~4 in order of increasing
$\Theta_1$. The first column shows the name of the galaxy,  the
second indicates its morphological type with detailed
classification of dwarf galaxies (Karachentsev et al.~2013). The
designations of parameters in the subsequent columns are the
same as in Table~2. The penultimate column shows the difference
between the line-of-sight velocities of the suite galaxy
and its MD. Some objects from the
list (KKR~8, KKH~65, KK~258, KK~227) coincide with the list of
isolated early-type galaxies in the Local Supercluster
(Karachentseva et al. 2010).

As we can see from Table~4, this list contains only the dwarf
systems with linear diameters of less than 4~kpc and absolute
magnitudes not brighter than   $-16.5^m$.  We have classified  a
half of them as transition objects (Tr) between dIr and dSph.
Three dwarf galaxies of S0 and E types: NGC~4600, NGC~404 and
NGC~59 reveal a gas content according to the optical emission 
spectra and HI fluxes. In fact, only 4 out of 12 galaxies: 
KKR~8, KKH~65, KKR~25 and UGC~8882 remain well founded representatives 
of isolated early-type galaxies. Moreover, only one of them, KKR~25
was studied in detail in the optical and radio ranges (Makarov et
al. 2012) and has a reliable distance estimate by the TRGB method
(Karachentsev et al. 2001).

From the aspect of evolution of dwarf galaxies,  of
great interest here are not only the isolated early-type galaxies,
but also the gas-rich dwarfs of Ir, Im, BCD types which are
located closely to the massive galaxies. They occupy the opposite
diagonal corner on the \{$T, \Theta_1$\} diagram  with respect to
the isolated early-type objects. Table~5 lists the data on 18
irregular dwarf galaxies,  T = 9,10 types, with the tidal indices
$\Theta_1> 3.0$  around the giant galaxies with absolute
magnitudes $M_B<-20.0^m$. The galaxies here are ranked according
to their  $\Theta_1$. The parameter designations in the
columns are the same as in Table 4.

As one can see, the majority of dwarf galaxies of this list are
detected in the HI line. This may assume that other yet
undetected dwarf systems have significant amounts of neutral
hydrogen, but they are too close to the massive galaxies and are
not resolved as individual HI-sources.

The average absolute magnitude of the dwarfs in Tables~4 and 5 is
almost identical: $-12.6^m$ and $-12.8^m$, respectively. This
agreement is to be expected if the late-type dwarf galaxies are
experiencing their first passage near the massive galaxy, and
after that, being deprived of their gas, move to the category of
spheroidal dwarfs.

Attention is drawn to an inhomogeneous distribution of the number
of irregular dwarfs, which are tightly located around the MDs.
This way, four dwarf galaxies are
close to the M81 and all of them are young stellar systems formed
in the tidal HI-filaments connecting M81 with M82 and
NGC~3077 (Yun 1999, Makarova et al. 2002, Karachentsev et al.
2011). Two giant spiral galaxies: NGC~6744 and NGC~6946 have 4 and
3 irregular dwarfs in their close vicinities, respectively. The
Milky Way and 6 other MDs have only one such companion each. (We
have not included the SMC galaxy in Table~5 because its MD is not
the Milky Way, but the LMC galaxy.) At the same time, such massive
galaxies as M31, Centaurus~A and Sombrero (NGC~4594) have no
nearby gas-rich dwarf companions at all.

It should be noted, however, that among the dwarf galaxies from
Table~5, only one galaxy --- LMC has its distance measured with
high accuracy. In the other objects of this list the distance
error is about 25\%.

As can be seen from the Table 5, the Milky Way stands out among the
other MDs by the presence of a nearby massive companion, LMC.
This peculiarity of the Milky Way was   noted by Rodriguez-Puebla
et al. (2013), Jiang et al. (2012) and other authors. This fact
remains valid if we not only  consider the T = 9, 10 dwarf
companions, but also all other types of companions. Around the 20
most significant MDs of the Local Volume (Table 2) there are 27
physical companions with $\Theta_1>0$ and   absolute magnitudes
brighter than $-17.0^m$. The distribution of these galaxies by
\{$\Theta_1, M_B$\}  is shown in Fig.~6. We have also placed
there the SMC galaxy, which lies in the potential well of the
Milky Way, although its MD is the LMC (see Table 1). As one can
see, some giant galaxies have physical companions of high
luminosity, such as M33 in M31, NGC~3351 in NGC~3368 and NGC~2835
in NGC~2784. However, they are not located as close to their MDs,
as the LMC to our Galaxy. Note that among the 27 massive nearby
companions in Fig.~6, all but NGC~3412 are late-type galaxies
with large amounts of neutral gas and active star formation.
This circumstance may indicate that many gas-rich companions are
still in the initial stage of falling towards their MDs.

As follows from Fig.~6, the Milky Way with the suite of its
companions does not look like a quite typical group. This
observational fact should be taken into account when comparing
the results of N-body simulations (Knebe et al. 2011,
Libeskind et al.2010) with the properties of the
galaxies in the  Local Group.

According to \{$\Theta_1, M_B$\} diagram the suite around M81 is 
most similar to our Galaxy with its neighbors.
However, the  M81 group has its
essential features: the presence of HI-filaments (Yun 1999),
young ``tidal'' Holm~IX- type dwarfs (Makarova et al. 2002), 
and also BCD galaxies (Chiboucas et al. 2009), which all are 
absent in the Local Group.

\section{On the kinematics of companions in MD suites}

The penultimate column of  Table~1 shows the line-of-sight
velocities of galaxies of the suite relative to the velocity of
the main galaxy. These data provide important
information about the kinematics and dynamics of the nearest
groups. The distribution of line-of-sight velocity difference in
20 most populated suites around their massive main
galaxies is shown in the left panels of Fig.~7. Physical members
of the groups with $\Theta_1>0$ are depicted by filled circles,
while the peripheral objects (or field galaxies) are
marked by open circles. As we can see from the top panel, the
variance of the line-of-sight velocity difference is almost
independent on the value of $\Theta_1$ in the region of
$\Theta_1>0$.

All the group members, except one lie in the strip of $\pm300$ km
s$^{-1}$. However, among the field galaxies with $\Theta_1<0$,
there are some cases with a large line-of-sight velocity difference,
for example, dwarf galaxies VCC~114 and VCC~1675 in front of the
Virgo cluster, for which the giant Sombrero galaxy (NGC~4594)
turned out to be the MD. Increasing relative velocity scatter in
the region of  $\Theta_ 1<0$ is quite expected and indicates the
absence of a physical relation of such galaxies with their MDs.

The bottom left panel of Fig.~7 compares the velocity difference
in the suite galaxies with the absolute magnitude of their main
galaxy. In the physical group members (filled circles), the
velocity dispersion tends to decrease towards the low luminosity
of main galaxies.

The right-hand panels of Fig.~7 represent the same data for the
least populated suites  which are composed of one galaxy only.
The luminosities and masses of   MDs with one companion are much
lower than those of the main galaxies of 20 populated suites.
Obviously, for this reason, the variation in the line-of-sight
velocity difference there lies in a narrower strip of only
$\pm200$ km s$^{-1}$, which is substantially lower than in the
companions of  massive galaxies.

It should be mentioned that a significant part of   galaxies in
the nearby suites has no line-of-sight velocity measurements to
date. Filling this gap is an urgent observational task.

\section{Concluding remarks}

The above data show that nearby groups of galaxies significantly
differ from each other in the structure and morphological
composition of its population. This fact should be taken into account
when comparing the results of N-body simulations of the
Cosmic Web structure with the observational data.

Usually, an object of such a comparison is the Local Group
(Libeskind et al. 2010, Zavala et al. 2009, Knebe et al. 2011),
which consists of two dynamically isolated suites of dwarf
galaxies around the Milky Way and Andromeda (M31), approaching
each other with the mutual velocity of centers
$\sim100 $km s$^{-1}$. However,  by a number of features the Local
Group is not typical among the nearby groups. Therefore, a
comparison of the results of numerical simulations should be
conducted with the characteristics of the mean (synthetic)
group of the Local Volume, relying in particular on the data of
Table 2.

One of important observational parameters of nearby  groups is
the number of their members. To characterize not a single group,
but rather their ensemble in a certain volume we usually chose
the value of the mean group population,   $\langle n_g\rangle$.
Trentham \& Tully (2002)   added another dimensionless variable
to this parameter, the ratio of the numbers of dwarf and normal
galaxies  $n_d/n_n$, which according to them varies greatly from
one group to another. It is easy to see that both of these
parameters,  $\langle n_g\rangle$ and $\langle n_g/n_n\rangle$
are not robust characteristics, as they are sensitive to the
choice of the threshold absolute magnitudes for dwarf galaxies
and normal ones.

If one considers the belonging a certain galaxy to its Main
Disturber as an analogue of a bibliographic reference, then the
ensemble of suites around the MDs in a fixed volume can be
described by a single number -- the h-index  suggested by
Jorge~E.~Hirsch (2005). The value of $h$ equal to, say, 10  means that
the given volume contains 10 suites (or groups) with the number of
companions to the MDs of 10 or more. According to the data of Table~1, 
we see that the suites in the Local Volume are characterize by h-index of
$n_s=13$. Ignoring the  suite members with $\Theta_1<0$ as the
general field galaxies, for the physical groups of
galaxies in the Local Volume we obtain the  $h_g=9$ index. 
The h-index is quite robust. 
If we exclude the ultra-dwarf galaxies with $M_B>-11.0^m$ from the
group members,  then the h-index for the groups will remain
unchanged.

We have to note that, in general, the suite can show hierarchy structure: 
the main suite can contains sub-suites. 
Table 1 gives us these examples:
the LMC belongs to the suite of the Milky Way, but at the same
time it is the MD for its close neighbor --- SMC; another example
in the Local Group is the dwarf spheroidal galaxy And XXII near M 33, which
is itself a member of the suite of M 31. In such cases, the
populations of  ``secondary'' sub-suites can be considered as members 
of the general suite around the most massive galaxy. Nevertheless,
it follows from the Table~1 data that the account of the presence
of hierarchical sub-groups does not change the value of the
h-index $h_g=9$ for the Local Volume.

The nowadays sky surveys in the optical range and in the
HI line discover new galaxies in the Local Volume and
measure/refine their line-of-sight velocities. The Hubble Space
Telescope  continues the programs measuring the distances to the
nearby galaxies. These targeted efforts promise to make soon 
the Local Volume as the fair sample for the analysis of various
properties of galaxies and their systems.

{\bf Acknowledgements}

This work was supported by the Russian Foundation for Basic
Research through grants RFBR-DFG 12--02--91338, RFBR-RUS-UKR 12--02--90407
and RFBR 11-02-00639, as well as the Ministry of Education and Science of
the Russian Federation (project 8523). Support for the HST proposals 
GO 12546, 12877, and 12878 was provided by NASA through grants from the 
Space Telescope Science Institute. We acknowledge the usage of the 
HyperLeda database (http://leda.univ-lyon1.fr). This research has made use 
of the NASA/IPAC Extragalactic Database (NED), which is operated by the Jet 
Propulsion Laboratory, California Institute of Technology, under contract 
with the National Aeronautics and Space Administration.

{\bf References}

Bell E.F., McIntosh D.H., Katz N., Weinberg M.D., 2003, ApJS, 149, 289

Benitez-Llambay A., Navarro J.F., Abadi M.G. et al., 2013, ApJ, 763L, 41

Chiboucas K., Karachentsev, I.D., Tully R.B., 2009, AJ, 137, 3009

Hirsch J.E., 2005, Proceedings of the National Academy of Science, 102, 16569 (arXiv:physics/0508025)

Jiang C.Y., Jing Y.P., Li Cheng, 2012, arXiv:1209.5930

Jones D.H., Peterson B.A., Colless M., Saunders,W., 2006, MNRAS, 369, 25

Kaisina E.I., Makarov D.I., Karachentsev, I.D., Kaisin S.S., 2012, AstBu, 67, 115

Karachentsev I.D., Makarov D.I., 1999, IAU Symposium, 186, 109

Karachentsev I.D., Makarov D.I., Kaisina E.I., 2013, AJ, 145, 101

Karachentsev I.D., Kaisina E.I., Kaisin S.S., Makarova L.N., 2011, MNRAS, 415L, 31

Karachentsev, I.D., 2005, AJ, 129, 178

Karachentsev I.D., Kasparova A.V., 2005, Astronomy Letters, 31, 152

Karachentsev I.D., Karachentseva V.E., Huchtmeier W.K., Makarov D.I., 2004, AJ, 127, 2031

Karachentsev I.D., Sharina M.E., Dolphin A., et al, 2001, A \& A, 379, 407

Karachentseva V.E., Karachentsev I.D., Sharina M.E., 2010, Astrophysics, 53, 513 (arXiv:1104.2506)

Kitaura F.S., Erdogdu P., Nuza S.E. et al., 2012, MNRAS, 427L, 35

Klypin A., Hoffman E., Kravtsov A., Gottloeber S., 2003, ApJ, 596, 19

Klypin A., Trujillo-Comez S., Primack J., 2011, ApJ, 740, 102

Knebe A., Libeskind N.I., Doumler T., et al. 2011, MNRAS, 417L, 56

Lee M.G., Freedman W.L., \& Madore B.F., 1993, AJ, 106, 964

Libeskind N.I., Yepes G., Knebe A., et al, 2010, MNRAS, 401, 1889

Makarov D.I., Uklein R.I., 2012, Astr. Bull., 67, 135

Makarov D.I., Makarova L.N., Sharina M.E. et al., 2012, MNRAS, 425, 709

Makarov D.I., Karachentsev I.D., 2011, MNRAS, 412, 2498

Makarova L.N., Grebel E.K., Karachentsev I.D., et al, 2002, A \& A, 396, 473

Rodrigues-Puebla A., Avila-Reese V., Drory N., 2013, ApJ, 773, 172

Slater C.T., Bell E.F., 2013, ApJ, 773, 17

Trentham N., Tully R.B., 2002, MNRAS, 335, 712

Tully R.B., Shaya E.J., Karachentsev I.D. et al., 2008, ApJ, 676, 184

Yun M.S., 1999, in IAU symposium, vol.186, Galaxy Interactions at Low and High Redshift, J.E. Barnes \& D.B.Sanders, ed., p.81

Wang W., White S., 2012, MNRAS, 424, 2574

Zavala J., Jing Y.P., Faltenbacher A., et al, 2009, ApJ, 700, 1779

%\end{document}
 \begin{table}
 \caption{LV galaxies ranked according to their Main Disturbers and Tidal Index.}
 \begin{tabular}{lrrrrrrcrrr} \hline
\hline
  Name             & $A_{26}$ & $M_B$    &$\lg M_*$   &$\lg M_{26}$ &$\lg M_{HI}$  & $\Theta_1$  & MD          &  $D$    & $\Delta v$  &$n_s$  \\
\hline
  HolmIX            & 2.96& $-$13.6  & 7.70  &  8.53&  8.40 &  5.1 &MESSIER081   &  3.61 &   88 & 53  \\
  ClumpI            & 0.20&  $-$8.3  & 5.57  &      &       &  4.2 &MESSIER081   &  3.60 & $-$129 & 53  \\
  KDG061            & 1.55& $-$12.9  & 8.09  &      &       &  4.0 &MESSIER081   &  3.60 &  256 & 53  \\
  $[$CKT2009$]$d0959+68 & 0.88& $-$10.1  & 6.29  &      &       &  4.0 &MESSIER081   &  3.60 & $-$150 & 53  \\
  ClumpIII          & 0.11&  $-$8.3  & 5.57  &      &       &  3.9 &MESSIER081   &  3.60 &  $-$85 & 53  \\
  NGC2976           & 6.17& $-$17.1  & 9.42  &  9.15&  8.03 &  2.9 &MESSIER081   &  3.56 &   38 & 53  \\
  MESSIER082        &13.16& $-$19.6  &10.57  &  9.86&  8.95 &  2.8 &MESSIER081   &  3.53 &  224 & 53  \\
  KDG064            & 2.19& $-$12.6  & 7.98  &      &       &  2.7 &MESSIER081   &  3.70 &   17 & 53  \\
  $[$CKT2009$]$d0934+70 & 0.87&  $-$9.6  & 6.80  &      &       &  2.5 &MESSIER081   &  3.66 &      & 53  \\
  IKN               & 3.15& $-$11.6  & 7.60  &      & $<$6.20 &  2.5 &MESSIER081   &  3.75 & $-$105 & 53  \\
  HIJASS J1021+6842 &     &        &       &      &  7.51 &  2.3 &MESSIER081   &  3.70 &   83 & 53  \\
  $[$CKT2009$]$d0939+71 & 0.38&  $-$8.4  & 5.60  &      &       &  2.3 &MESSIER081   &  3.63 &      & 53  \\
  KK77              & 2.89& $-$12.0  & 7.76  &      & $<$6.14 &  2.2 &MESSIER081   &  3.48 &      & 53  \\
  F8D1              & 2.70& $-$12.6  & 7.99  &      & $<$6.20 &  2.2 &MESSIER081   &  3.77 &  $-$96 & 53  \\
  KDG063            & 2.18& $-$12.1  & 7.80  &  6.70&  6.91 &  2.0 &MESSIER081   &  3.50 & -104 & 53  \\
  DDO078            & 2.23& $-$11.5  & 7.54  &      &       &  1.9 &MESSIER081   &  3.72 &   87 & 53  \\
 HolmI             & 5.29& $-$14.5  & 8.01  &  7.95&  8.01 &  1.7 &MESSIER081   &  3.84 &  187 & 53  \\
 $[$CKT2009$]$d1006+67 & 0.39&  $-$8.5  & 6.37  &      &       &  1.6 &MESSIER081   &  3.87 &      & 53  \\
 $[$CKT2009$]$d0955+70 & 0.48&  $-$9.1  & 6.60  &      &       &  1.5 &MESSIER081   &  3.93 &      & 53  \\
 KDG073            & 1.29& $-$10.8  & 6.56  &  6.39&  6.51 &  1.4 &MESSIER081   &  3.70 &  159 & 53  \\
 UGC05497          & 0.96& $-$12.3  & 7.18  &  6.11&  6.02 &  1.4 &MESSIER081   &  3.70 &  163 & 53  \\
 $[$CKT2009$]$d0926+70 & 0.56& $-$10.0  & 6.24  &      & $<$5.64 &  1.3 &MESSIER081   &  3.93 &      & 53  \\
 $[$CKT2009$]$d0944+69 & 0.26&  $-$7.4  & 5.92  &      &       &  1.3 &MESSIER081   &  3.98 &      & 53  \\
 BK6N              & 1.23& $-$11.1  & 7.38  &      &       &  1.2 &MESSIER081   &  3.85 &      & 53  \\
 HS117             & 1.90& $-$11.2  & 6.72  &  6.25&  5.01 &  1.2 &MESSIER081   &  3.96 &   12 & 53  \\
 BK3N              & 0.40&  $-$9.6  & 6.09  &  5.82&  7.26 &  1.2 &MESSIER081   &  4.02 &   $-$3 & 53  \\
 \hline
 \multicolumn{11}{l}{\small{\textbf{Note.} Only a portion of this table is shown here to demonstrate its form and content.}}\\
\multicolumn{11}{l}{\small{     Machine-readable version of the full table is available.}}\\
\end{tabular}
\end{table}

\begin{table}
\caption{Properties of the 20  most populated suites in the LV.}
\begin{tabular}{lrrrrllllrrrrr} \hline
  MD   &   D   & $n_s$& $n_g$& $n_b$&   $M_B$  &  $\lg M_*$& $\lg M_{26}$ & $A_{26}$ &$\lg M_{HI}$ & $T$ & $\Theta_1$ & $\Theta_5$ & $\Theta_j$ \\
\hline
M.Way  &  0.01 & 38 & 29 &  5 & $-$20.8: & 10.5:& 11.3: & 25:  & 9.5: & 4 & 2.8 & 2.9 & 1.6 \\
M 31   &  0.77 & 42 & 39 & 10 & $-$21.40 & 10.73& 11.50 & 43.4 & 9.73 & 3 & 4.9 & 4.9 & 1.4 \\
IC 342 &  3.28 & 10 &  9 &  9 & $-$20.69 & 10.60& 11.15 & 34.2 &10.16 & 6 & 0.1 & 0.5 & 1.7 \\
M 81   &  3,63 & 53 & 37 & 22 & $-$20.92 & 10.93& 11.27 & 31.4 & 9.44 & 3 & 2.5 & 2.6 & 1.5 \\
N 5128 &  3.75 & 37 & 26 & 16 & $-$20.78 & 10.91& 11.70 & 42.6 & 8.46 &-2 & 0.7 & 1.0 & 1.6 \\
N 253  &  3.94 & 25 &  8 &  7 & $-$21.29 & 11.04& 11.24 & 40.8 & 9.15 & 5 &$-$0.4 &$-$0.3 & 0.7 \\
N 4826 &  4.37 & 11 &  3 &  3 & $-$19.51 & 10.48& 10.70 & 17.8 & 8.26 & 2 &$-$0.8 &$-$0.5 &$-$1.0 \\
N 4736 &  4.66 & 31 & 15 & 12 & $-$19.86 & 10.61& 10.73 & 20.7 & 8.32 & 2 &$-$0.6 &$-$0.1 & 0.8 \\
N 5236 &  4.92 & 28 & 15 & 14 & $-$20.64 & 10.86& 11.32 & 28.2 &10.00 & 5 &$-$0.5 & 0.0 & 0.0 \\
M 101  &  7.38 & 11 &  6 &  5 & $-$21.12 & 10.85& 11.35 & 65.2 & 9.91 & 6 & 0.4 & 0.5 & 0.2 \\
N 4631 &  7.38 & 16 &  5 &  4 & $-$20.28 & 10.49& 10.41 & 33.7 & 9.72 & 7 & 1.8 & 1.9 & 1.0 \\
N 2683 &  7.73 & 13 &  2 &  2 & $-$20.36 & 10.60& 11.14 & 29.5 & 8.94 & 3 & 0.0 & 0.2 &$-$1.3 \\
N 4258 &  7.83 & 31 & 19 & 17 & $-$21.20 & 10.94& 11.33 & 41.5 & 9.64 & 4 & 1.1 & 1.3 & 0.6 \\
N 6744 &  8.30 & 12 &  6 &  6 & $-$20.96 & 10.79& 11.35 & 52.8 &10.19 & 4 & 2.0 & 2.0 & 1.2 \\
N 2903 &  8.87 & 15 &  4 &  4 & $-$20.89 & 10.82& 11.13 & 32.4 & 9.44 & 4 & 1.7 & 1.7 &$-$0.8 \\
N 5055 &  8.99 & 11 &  5 &  5 & $-$20.98 & 10.99& 11.34 & 42.2 & 9.62 & 4 &$-$0.1 & 0.1 &$-$0.9 \\
N 4594 &  9.30 & 32 & 10 & 10 & $-$21.82 & 11.30& 11.76 & 32.5 & 8.36 & 1 & 2.5 & 2.6 &$-$0.4 \\
N 3115 &  9.68 & 12 &  7 &  7 & $-$20.77 & 10.95& 10.50 & 24.0 & 8.75 &$-$1 & 2.3 & 2.6 & 0.2 \\
N 2784 &  9.82 &  9 &  6 &  6 & $-$19.65 & 10.80&   -   & 19.3 & 8.0  &$-$2 & 3.1 & 3.2 & 1.0 \\
N 3368 & 10.42 & 31 & 31 & 31 & $-$20.40 & 10.83& 11.14 & 27.2 & 9.18 & 3 & 1.1 & 1.5 & 2.1 \\
\hline
\end{tabular}
\end{table}

\begin{table}
\caption{Correlation coefficients for the 20 most populated suites.}
\begin{tabular}{c|rccccrcccc} \hline
      &   $D$  &  $M_B$ & $\lg M_*$ & $\lg M_{26}$ & $\lg A_{26}$ & $\lg M_{HI}$ &  $T$   & $\Theta_1$  & $\Theta_5$  &$\Theta _j$ \\
\hline
 $n_s$  &--0.48 &--0.37 & 0.23 & 0.41  & 0.03 & --0.02 &--0.06 & 0.29 & 0.27 & 0.44  \\

 $n_g$  &--0.48 &--0.31 & 0.17 & 0.39  & 0.08 &  0.12  &--0.11 & 0.43 & 0.39 & 0.74    \\

 $n_b$  &--0.08 &--0.24 & 0.42 & 0.33  & 0.07  &  0.05  &-0.17  & 0.17 & 0.12 & 0.63      \\
\hline
\end{tabular}
\end{table}

\begin{table}
\caption{Early-type dwarfs with negative $\Theta_1$.}
\begin{tabular}{llcrrrrrrlrrr}
\hline
Name    &  Type  & $A_{26}$  &  $M_B$ &  $\lg M_*$ &  $\lg M_{26}$& $\lg M_{HI} $ & $\Theta_1$&   MD     &   D   &  $\Delta v$ & $n_s$ \\
\hline
KKR08   &  Sph--L & 1.35 &$-$11.7 &  7.62 &       & $<$7.02  &$-$1.6& NGC4594  &  8.00 &      & 32  \\
KKH65   &  Sph--L & 1.70 &$-$12.7 &  8.02 &       &          &$-$1.4& NGC3627  & 10.00 &      &  8  \\
UGC1703 &  Tr-- N & 1.05 &$-$11.5 &  7.56 &       & $<$6.30  &$-$1.3& Maffei2  &  4.19 &      &  8  \\
KDG216  &  Tr-- L & 1.99 &$-$12.1 &  7.78 &       & $<$6.77  &$-$1.2& NGC4826  &  6.00 &      & 11  \\
KK258   &  Tr-- L & 0.98 &$-$10.3 &  7.06 &       & $<$5.65  &$-$1.1& NGC0253  &  2.00 &      & 25  \\
KDG218  &  Tr-- L & 2.73 &$-$11.9 &  7.71 &       & $<$6.61  &$-$1.0& NGC5236  &  5.00 &      & 28  \\
KKR25   &  Sph--L & 0.59 &$ $-9.4 &  6.71 &       & $<$4.91  &$-$1.0& M 31     &  1.86 &  157 & 42  \\
NGC4600 &  S0e--N & 3.49 &$-$15.8 &  9.12 &       & $<$7.06  &$-$1.0& NGC4594  &  7.35 & $-$181 & 32  \\
NGC0404 &  S0e--N & 3.25 &$-$16.5 &  9.28 &   9.58& $ $7.93  &$-$0.8& Maffei2  &  3.05 &  $-$21 &  8  \\
KK227   &  Tr-- L & 2.06 &$-$12.5 &  7.97 &       &          &$-$0.6& NGC5055  & 10.00 &      & 11  \\
NGC0059 &  dEe--N & 3.52 &$-$15.7 &  8.72 &   8.25& $ $7.40  &$-$0.5& NGC0253  &  5.30 &  155 & 25  \\
Tucana  &  Tr-- L & 0.73 & $-$9.2 &  6.62 &       & $<$4.18  &$-$0.2& Milky Way&  0.88 &  138 & 38  \\
UGC8882 &  dE-- N & 2.58 &$-$13.9 &  7.67 &       &        & 0.0& M 101    &  8.32 &  104 & 11  \\
\hline
\end{tabular}
\end{table}

\begin{table}
\caption{Late-type T=9, 10 dwarfs with $\Theta_1>3.0$ around the
MDs brighter than $M_B=-20.0$.}
\begin{tabular}{llcrrrrrrlrr}
\hline
Name      & Type  &  $A_{26}$ & $M_B$   & $\lg M_*$  & $\lg M_{26}$ &$\lg M_{HI}$  & $\Theta_1$&   MD     &  D   &  $\Delta v$ & $n_s$ \\
\hline
HolmIX    & Ir-- N & 2.96 &$-$13.6  & 7.70  &  8.53 & 8.40  & 5.1& M 81     & 3.61 &   88 & 53  \\
$[$KK2000$]$71& Ir-- N & 4.41 &$-$14.7  & 8.13  &       &       & 4.7& NGC6744  & 8.30 &      & 12  \\
ClumpI    & Ir-- N & 0.20 & $-$8.3  & 5.57  &       &       & 4.2& M 81     & 3.60 & $-$129 & 53  \\
CKT0959+68& Ir-- L & 0.88 &$-$10.1  & 6.29  &       &       & 4.0& M 81     & 3.60 & $-$150 & 53  \\
$[$KK2000$]$72& Ir-- L & 1.36 &$-$11.9  & 7.00  &       &       & 4.0& NGC6744  & 8.30 &      & 12  \\
ClumpIII  & Ir-- N & 0.11 & $-$8.3  & 5.57  &       &       & 3.9& M 81     & 3.60 &  $-$85 & 53  \\
KKSG18    & BCD--N & 4.45 &$-$16.6  & 9.27  &       &       & 3.9& NGC3115  & 9.70 &   17 & 12  \\
KKSG20    & Ir-- N & 1.68 &$-$12.8  & 7.37  &  6.69 & 6.18  & 3.9& NGC3521  &10.70 &   38 &  4   \\
$[$KK2000$]$70& Ir-- L & 1.37 &$-$12.1  & 7.09  &       &       & 3.8& NGC6744  & 8.30 &      & 12   \\
LV1217+47 & Tr-- L & 0.69 &$-$11.0  & 6.66  &       &$<$6.44  & 3.6& NGC4258  & 7.80 &      & 31  \\
LMC       & Im-- N &10.06 &$-$17.9  & 9.42  &  9.44 & 8.66  & 3.5& M.Way    & 0.05 &   93 & 38  \\
ESO104-044& Ir-- L & 4.37 &$-$14.8  & 8.17  &  8.81 & 8.33  & 3.5& NGC6744  & 8.30 &  $-$92 & 12  \\
KK251     & Ir-- L & 3.48 &$-$13.6  & 7.70  &  8.32 & 8.05  & 3.5& NGC6946  & 5.89 &   78 &  8  \\
N2903-HI-1& Ir-- N & 0.71 &$-$11.7  & 6.92  &  5.99 & 6.42  & 3.3& NGC2903  & 8.90 &   27 & 15  \\
KK69      & Ir-- L & 3.15 &$-$12.2  & 7.12  &  6.65 & 7.51  & 3.3& NGC2683  & 7.70 &   53 & 13  \\
UGC11583  & Ir-- L & 4.71 &$-$14.3  & 7.98  &  8.98 & 8.27  & 3.3& NGC6946  & 5.89 &   74 &  8  \\
LeG13     & Ir-- N & 1.25 &$-$12.8  & 7.35  &  5.97 & 6.75  & 3.1& NGC3368  &10.40 &  $-$22 & 31   \\
KK252     & Ir-- L & 2.33 &$-$14.1  & 7.89  &  8.73 & 7.04  & 3.1& NGC6946  & 5.89 &   86 &  8   \\
\hline
\end{tabular}
\end{table}

\begin{figure}[p]
\includegraphics[scale=1.5]{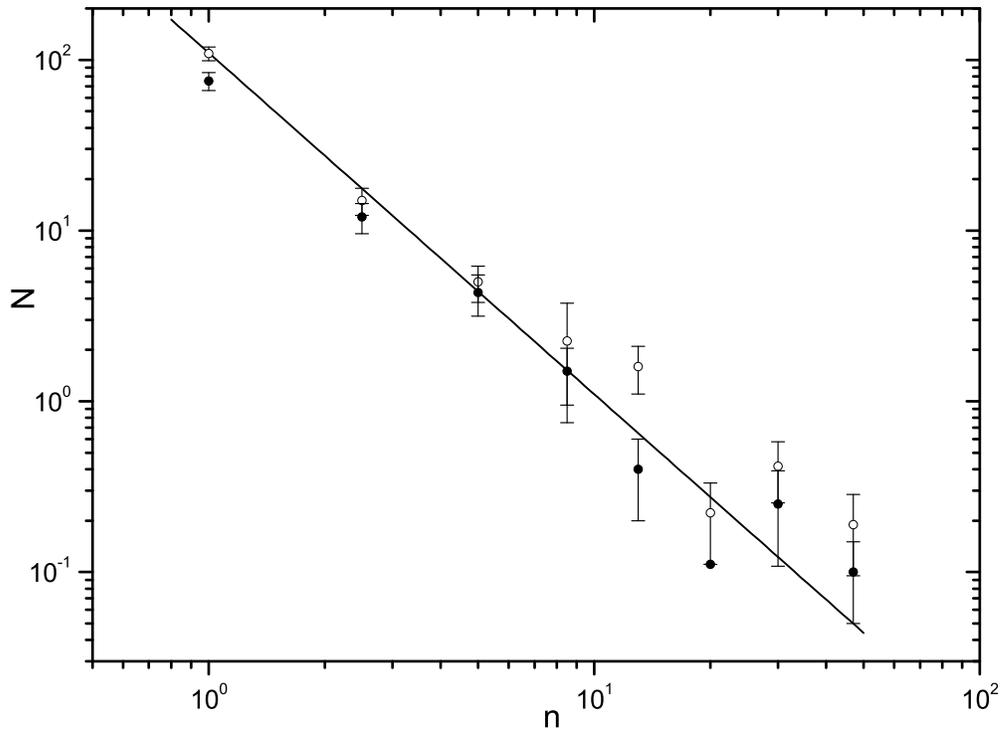}
\caption{The number of suites in the Local
Volume depending on the number of suite members (open circles)
and the number of dynamically bound companions with
$\Theta_1>0$ (filled circles). The straight line corresponds to the
$N(n)\propto n^{-2}$ relation.}
\end{figure}

\begin{figure}[p]
\includegraphics[scale=1.5]{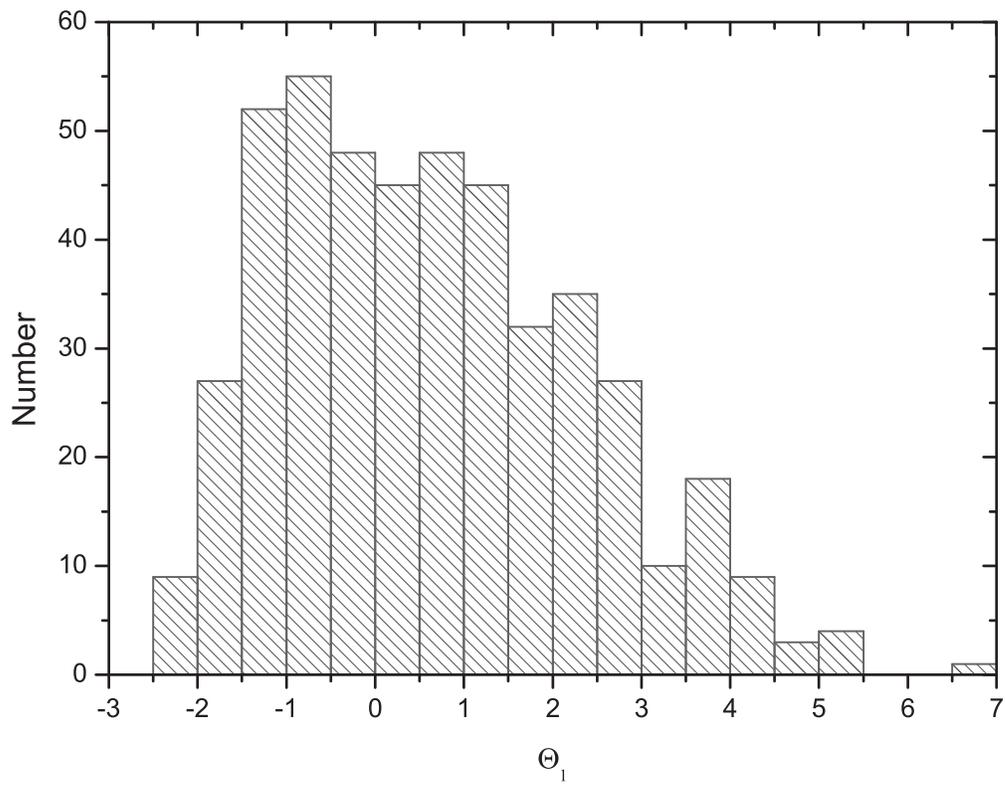}
\caption{Distribution of 468 galaxies in
the 20 most populated suites of the Local Volume by their
tidal index $\Theta_1$.}
\end{figure}

\begin{figure}[p]
\includegraphics[scale=1.5]{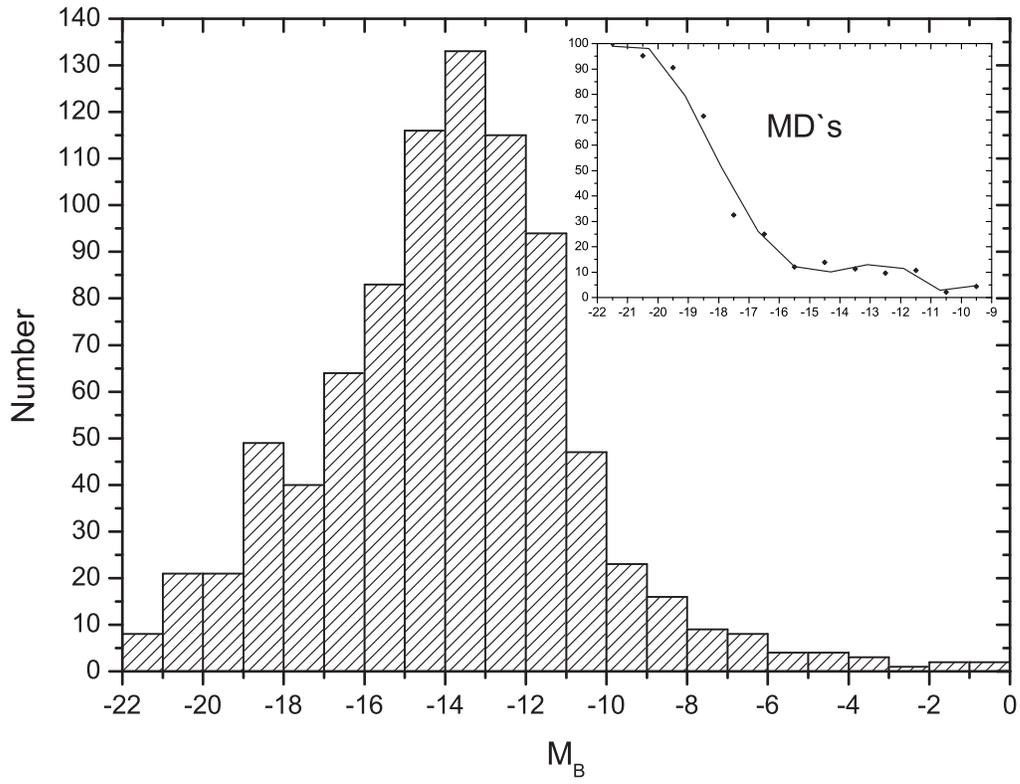}
 \caption{Distribution of 794 galaxies in a
sphere of  11 Mpc radius around the Milky Way on absolute
$B$-magnitudes, corrected for the internal and external
extinction. The inset shows what percentage of these galaxies in
each bin act as the Main Disturber.}
\end{figure}

\begin{figure}[p]
\includegraphics[scale=0.85]{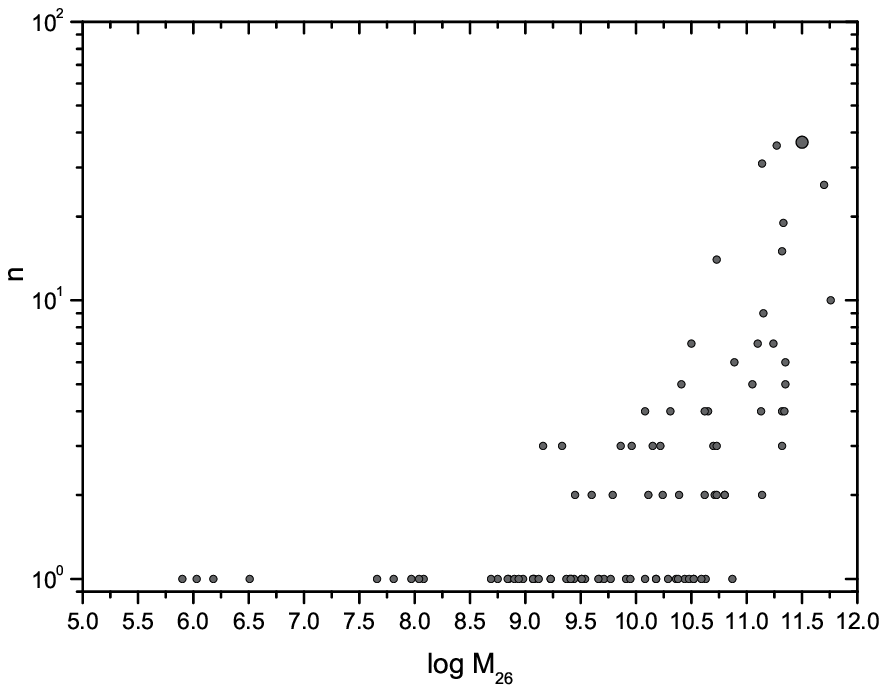}
\includegraphics[scale=0.85]{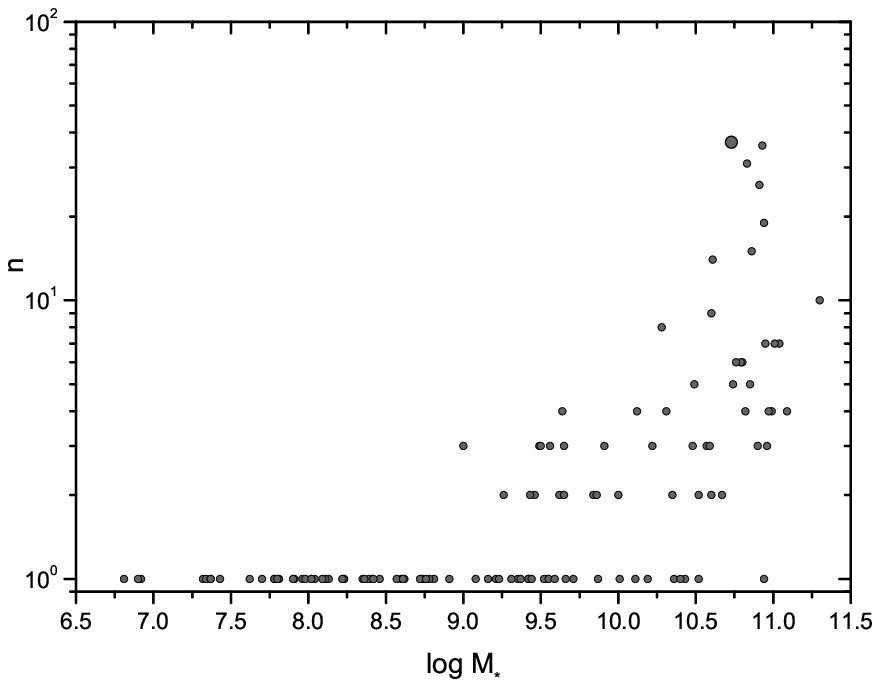}
\includegraphics[scale=0.85]{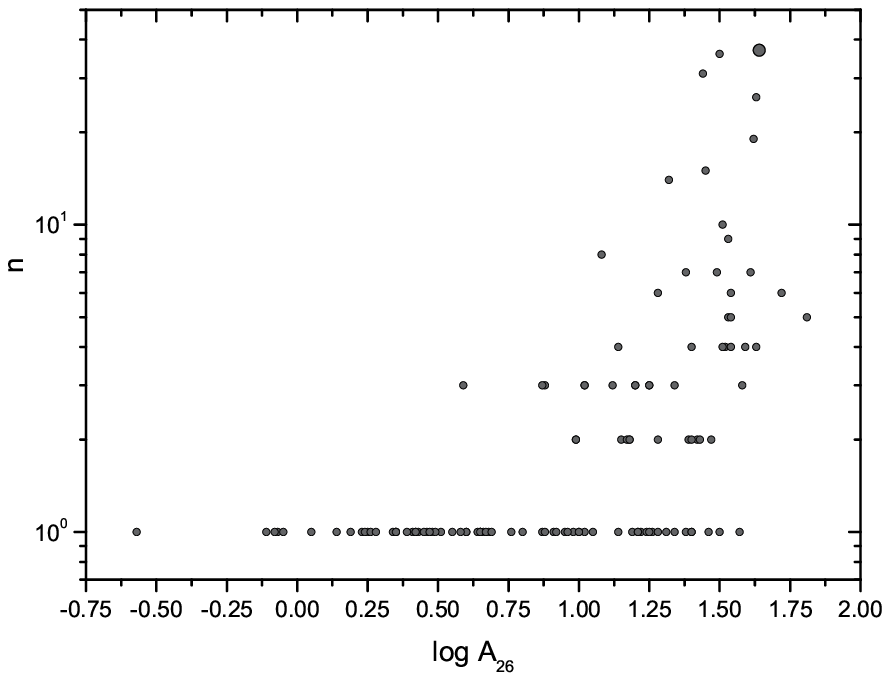}
\includegraphics[scale=0.85]{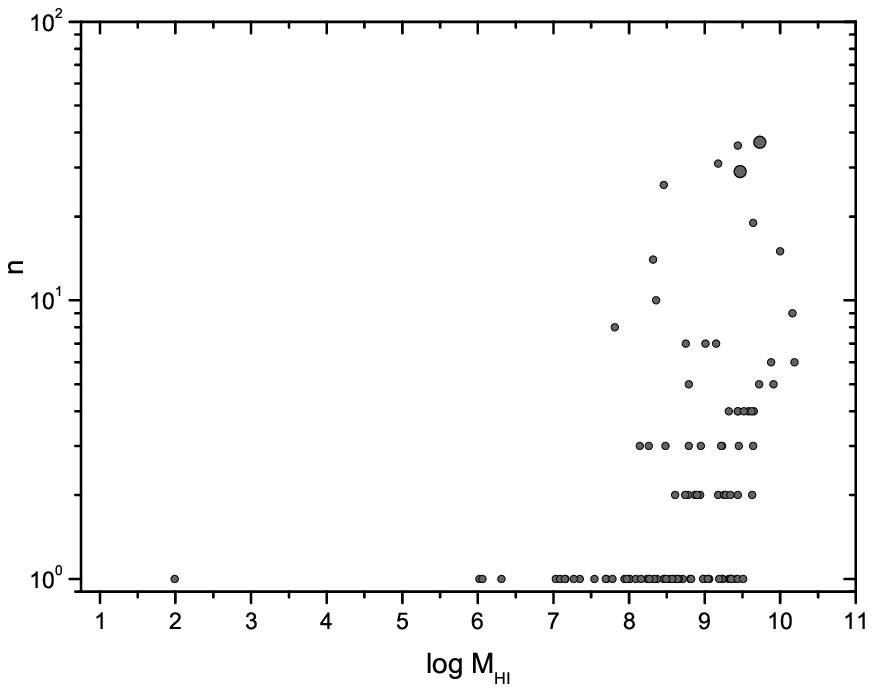}
 \caption{The number of physical companions
in the  Main Disturber suites as a function of MD global
parameters: dynamic mass $M_{26}$, stellar mass $M_*$, linear
diameter $A_{26}$, and hydrogen mass $M_{HI}$. The Milky Way
and M 31 suites are depicted by larger symbols.}
\end{figure}

\begin{figure}[p]
\includegraphics[scale=0.85]{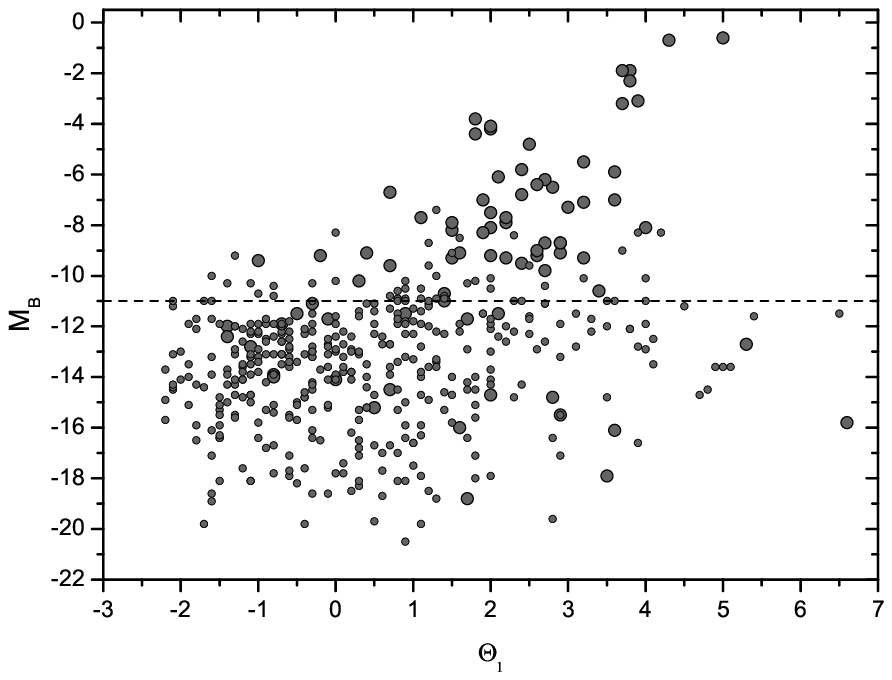}
\includegraphics[scale=0.85]{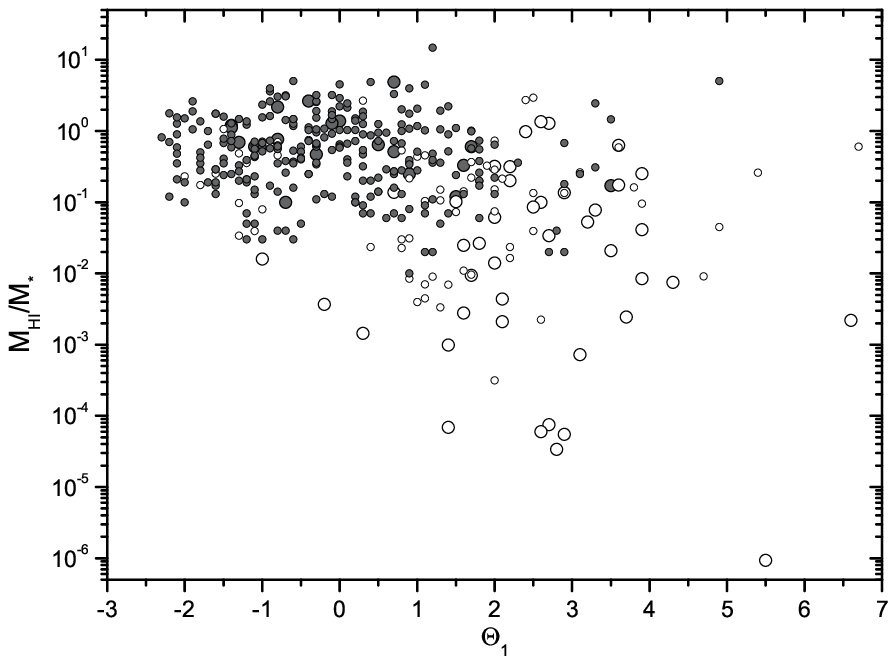}
\includegraphics[scale=0.85]{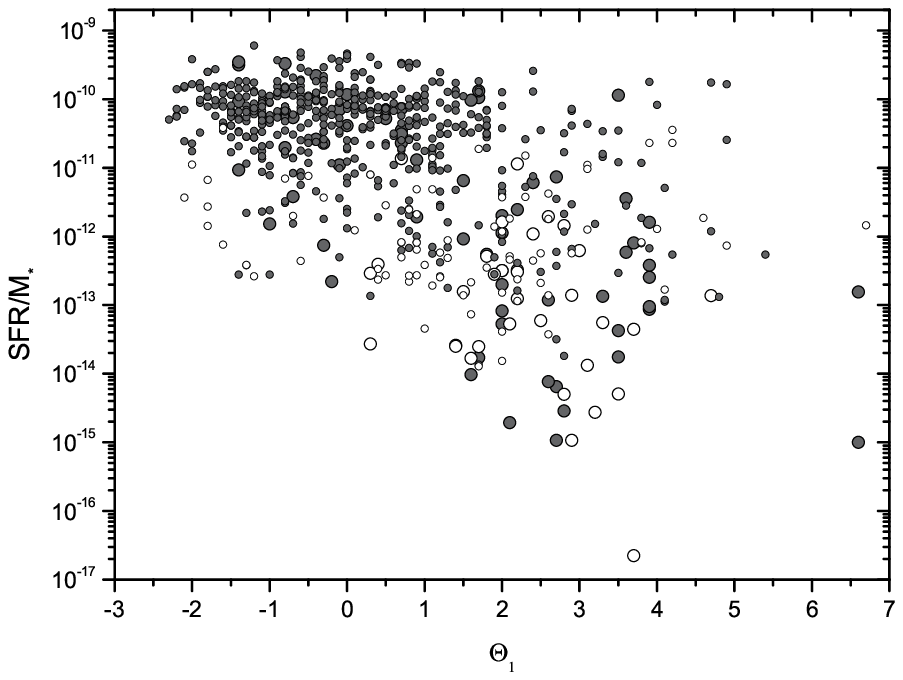}
\includegraphics[scale=0.85]{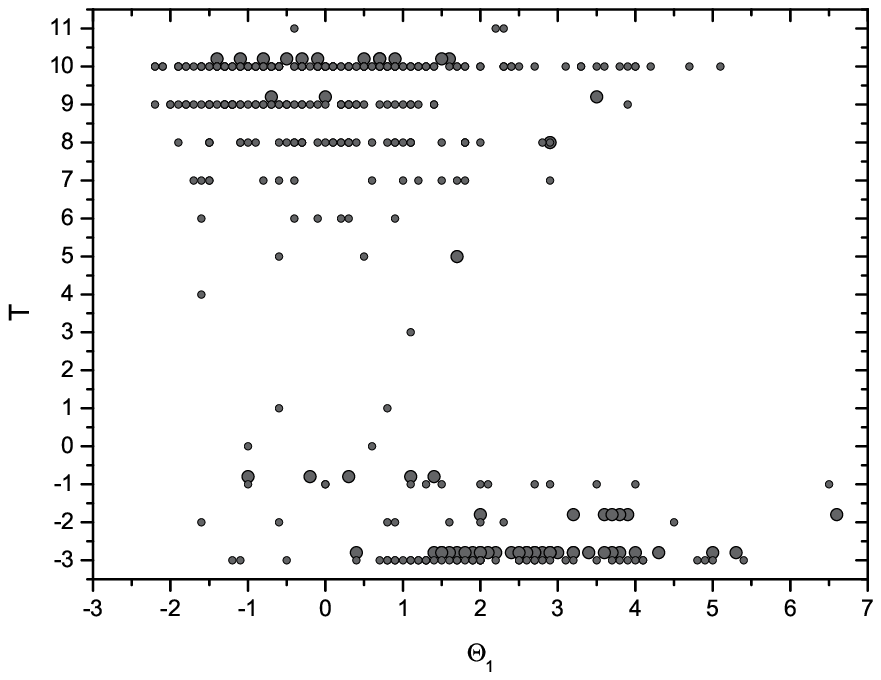}
\caption{Global parameters of galaxies in
20 most populated suites:  absolute magnitude   $M_B$,
hydrogen-to-stellar mass ratio $M_{HI}/M_*$, specific star
formation rate  $SFR/M_*$ and morphological type T by de
Vaucouleurs scale, depending on their tidal index $\Theta_1$.
Galaxies with the upper limit of $M_{HI}$ or $SFR$ are marked
with open symbols. The LG members are highlighted by larger
symbols.}
\end{figure}

\begin{figure}[p]
\includegraphics[scale=1.5]{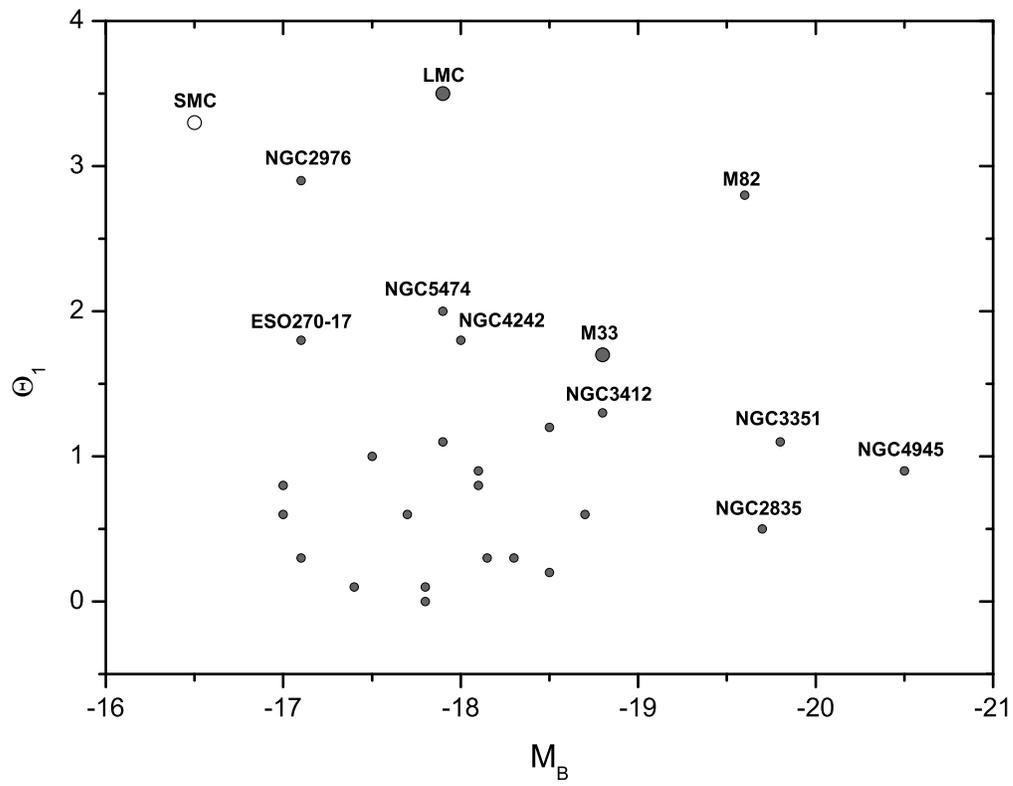}
 \caption{The distribution of physical
companions around  20 most massive galaxies of the Local Volume
by their tidal index and absolute magnitude.}
\end{figure}

\begin{figure}[p]
\includegraphics[scale=0.85]{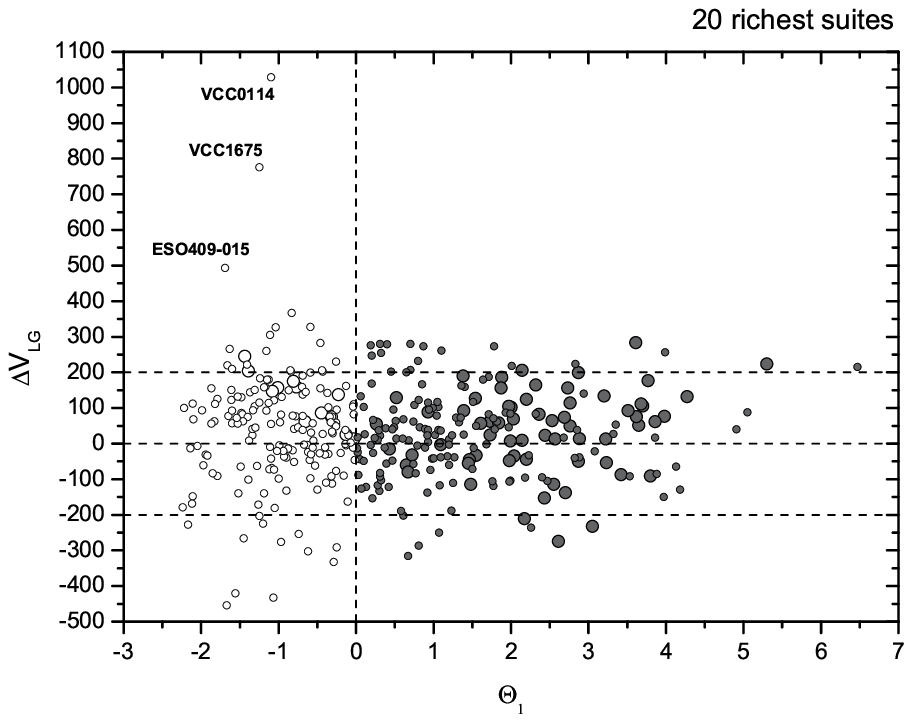}
\includegraphics[scale=0.85]{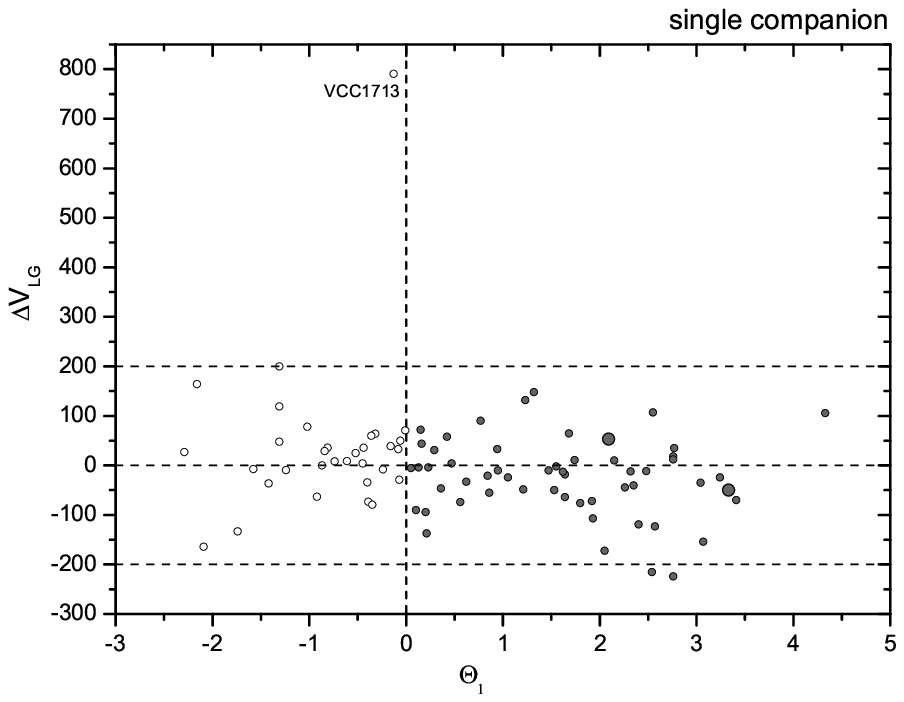}
\includegraphics[scale=0.85]{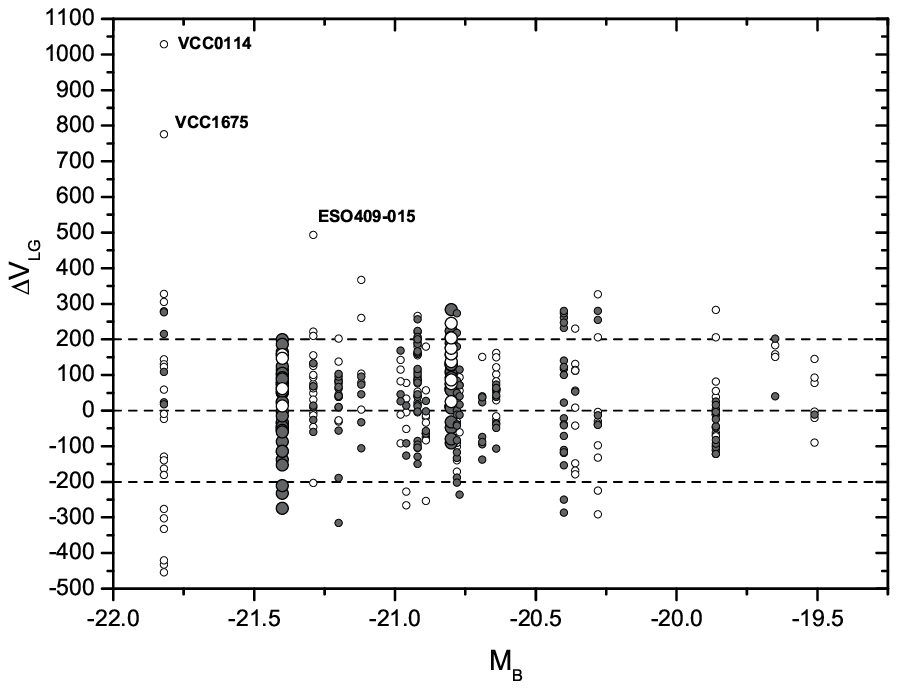}
\includegraphics[scale=0.85]{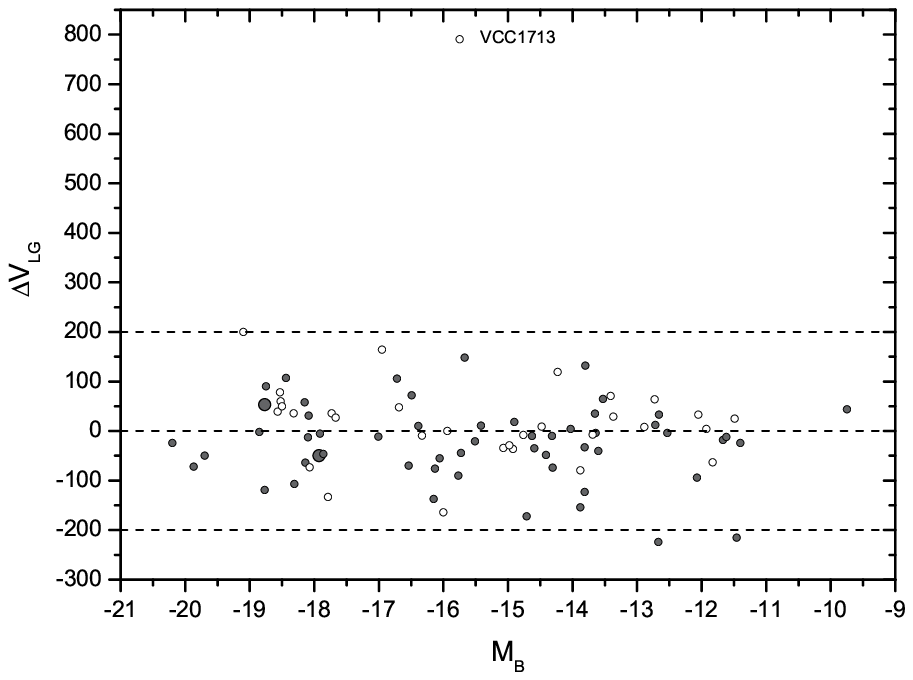}
 \caption{The line-of-sight velocity of the
suite member relative to its main galaxy as a function of the
suite member tidal index $\Theta_1$ and   absolute magnitude of
the main galaxy. The left panels correspond to the population
of   20 richest suites, right panels --- to the suites,
consisting of one companion. Physical companions with
$\Theta_1>0$  and field galaxies   ($\Theta_1<0$) are marked with
filled and open circles, respectively. The LG members are depicted
by larger circles.}
\end{figure}

\end{document}